\documentclass[3p,times]{elsarticle}

\usepackage{ecrc}
\usepackage{xspace}

\newcommand{\pt}{\mbox{$p_{T}$}\xspace}

\newcommand{\raa}{\mbox{$R_{\rm AA}$}\xspace}

\newcommand{\Npart}{\mbox{$N_{\rm part}$}\xspace}

\newcommand{\dndeta}{\mbox{$dN_{\rm ch}/d\eta$}\xspace}
\newcommand{\sloss}{\mbox{$S_{\rm loss}$}\xspace}

\newcommand{\piz}{\mbox{$\pi^0$}\xspace}

\volume{00}

\firstpage{1}

\journalname{Nuclear Physics A}

\runauth{}

\jid{npa}

\jnltitlelogo{Nuclear Physics A}

\usepackage{amssymb}

\biboptions{square,comma,numbers,sort&compress}

\usepackage[figuresright]{rotating}

\begin{document}

\begin{frontmatter}



\dochead{}

\title{Detail study of the medium created in Au+Au collisions with high \pt probes by the PHENIX experiment at RHIC}


\author{Takao Sakaguchi, for the PHENIX Collaboration}

\address{Brookhaven National Laboratory, Upton, NY 11973, USA.}

\begin{abstract}
Recent results on high \pt identified hadrons in Au+Au collisions from
the PHENIX experiment are presented. The \raa for
\piz and $\eta$ are found to be very consistent. The second and
fourth order collective flow of \piz's have been measured and found
that $v_4/v_2^2$ is consistent with the one observed in lower \pt region.
Assuming the suppression of the \piz yield at highest \pt arises
from energy loss of partons, we found that the energy loss is $L^3$
dependent, where $L$ is the path length of the partons in the
medium. The $\delta\pt/\pt$'s of high \pt hadrons are computed
from 39\,GeV Au+Au over to 2.76\,TeV Pb+Pb, and found that they
vary by a factor of six. We have seen a smooth trend in $\delta \pt/\pt$
from RHIC energy to LHC energy when plotting against charged multiplicity
of the systems.
\end{abstract}


\end{frontmatter}


\section{Introduction}
\begin{figure}[h]
\begin{minipage}{65mm}
\centering
\includegraphics[width=6.5cm]{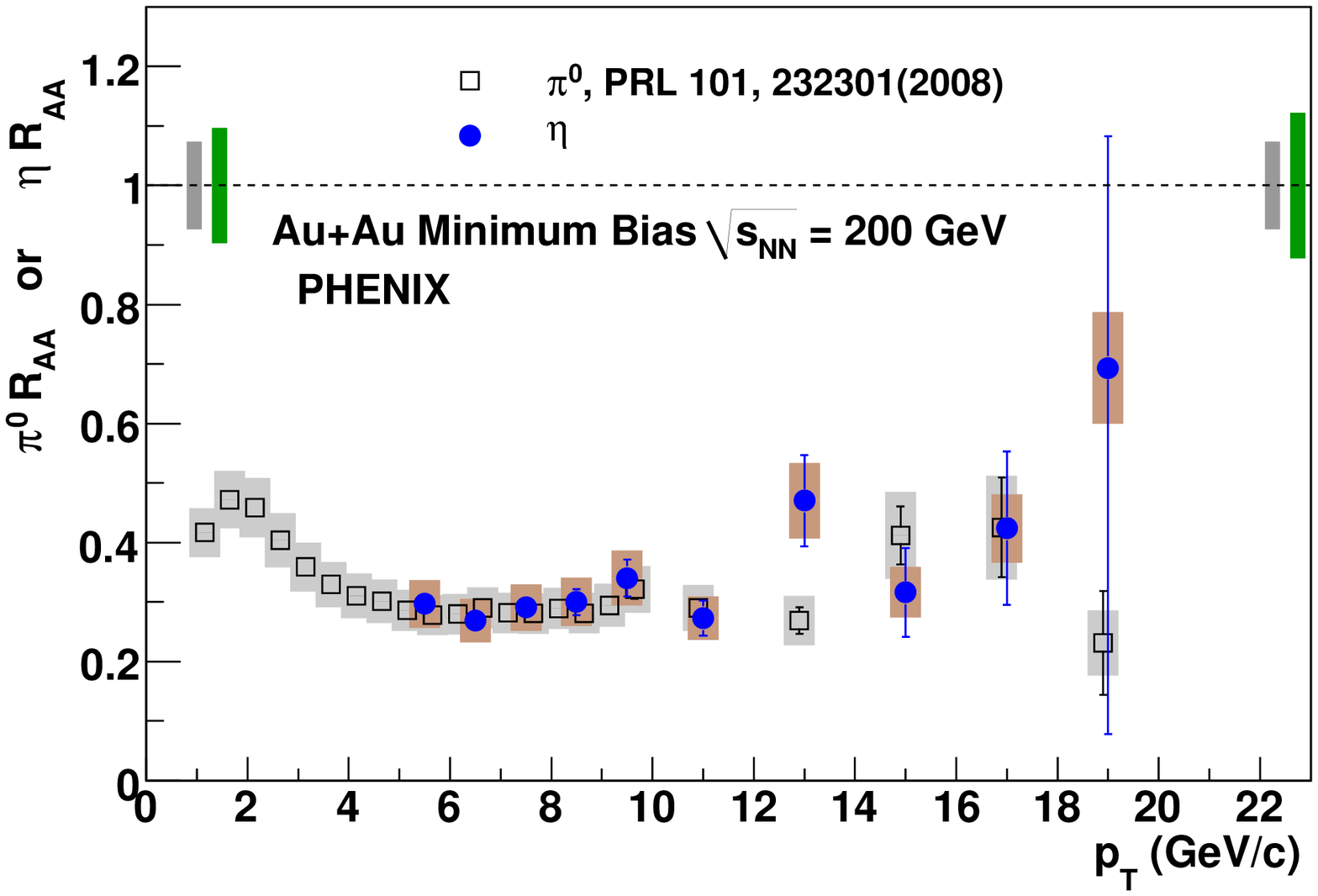}
\end{minipage}
\hspace{5mm}
\begin{minipage}{85mm}
\centering
\includegraphics[width=8.5cm]{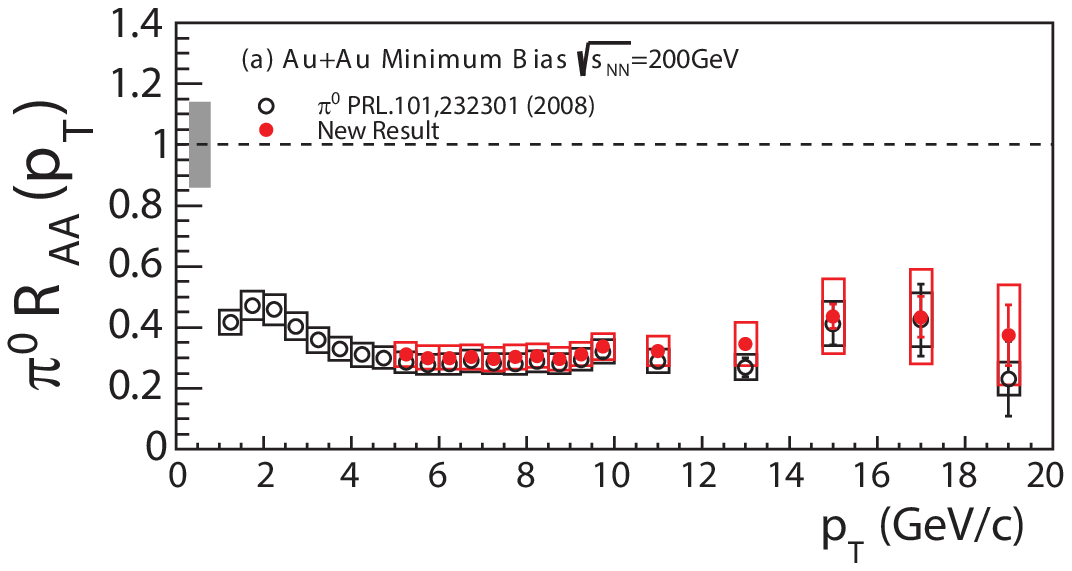}
\end{minipage}
\label{fig2}       
\caption{ (a, left) $R_{\rm AA}$ for $\pi^0$ and $\eta$ in minimum bias Au+Au
collisions. (b, right) $R_{\rm AA}$ for $\pi^0$ from the RHIC Year-2004 run
and Year-2007 run.}
\end{figure}

The interaction of hard scattered partons with the medium created by heavy ion
collisions (i.e., quark-gluon plasma, QGP) has been of interest since the
beginning of the RHIC running~\cite{Wang:1998bha}. A large suppression of the
yields of high transverse momentum (\pt) hadrons which are the fragments of
such partons was observed, suggesting that the matter is sufficiently dense
to cause parton-energy loss prior to hadronization~\cite{Adler:2003qi}.
The PHENIX experiment~\cite{Adcox:2003zm} has been exploring the highest
\pt region with single \piz and $\eta$ mesons, which are leading
hadrons of jets, and thus provide a good measure of momentum of hard
scattered partons. Here, we present the recent results obtained from Au+Au
collisions in the Year-2007 run (0.81 nb$^{-1}$).
Fig.~\ref{fig2}(a) shows the nuclear modification factors
\raa ($\equiv(dN_{\rm AA}/dydp_{T})/(\langle T_{\rm AA}\rangle d\sigma_{pp}/dydp_{T})$) for $\pi^0$ and $\eta$'s in 200\,GeV Au+Au
collisions~\cite{Adare:2010dc}.
They are very consistent each other in spite of hidden strangeness contents
in $\eta$ mesons. This also implies that the fragmentation function
is not modified by the medium for the $p_T$ range we measured. 
Because $\eta$ has four times larger mass compared to that of $\pi^0$,
one can resolve two photons decaying from $\eta$ up to four times
larger $p_T$ of $\pi^0$, resulting in a higher $p_T$ reach with smaller
systematic errors with $\eta$. The \raa's in Fig.~\ref{fig2}(b)
demonstrates that the $\pi^0$ yield from the Year-2007 run has smaller
errors and is consistent with that from the Year-2004 run~\cite{Adare:2012wg}.
We used the same $p+p$ reference for \raa's from both Au+Au running.

\section{Anisotropy of high $p_T$ $\pi^0$ yield}
\subsection{Collective flow}
\begin{figure}[h]
\begin{minipage}{80mm}
\centering
\includegraphics[width=7.5cm]{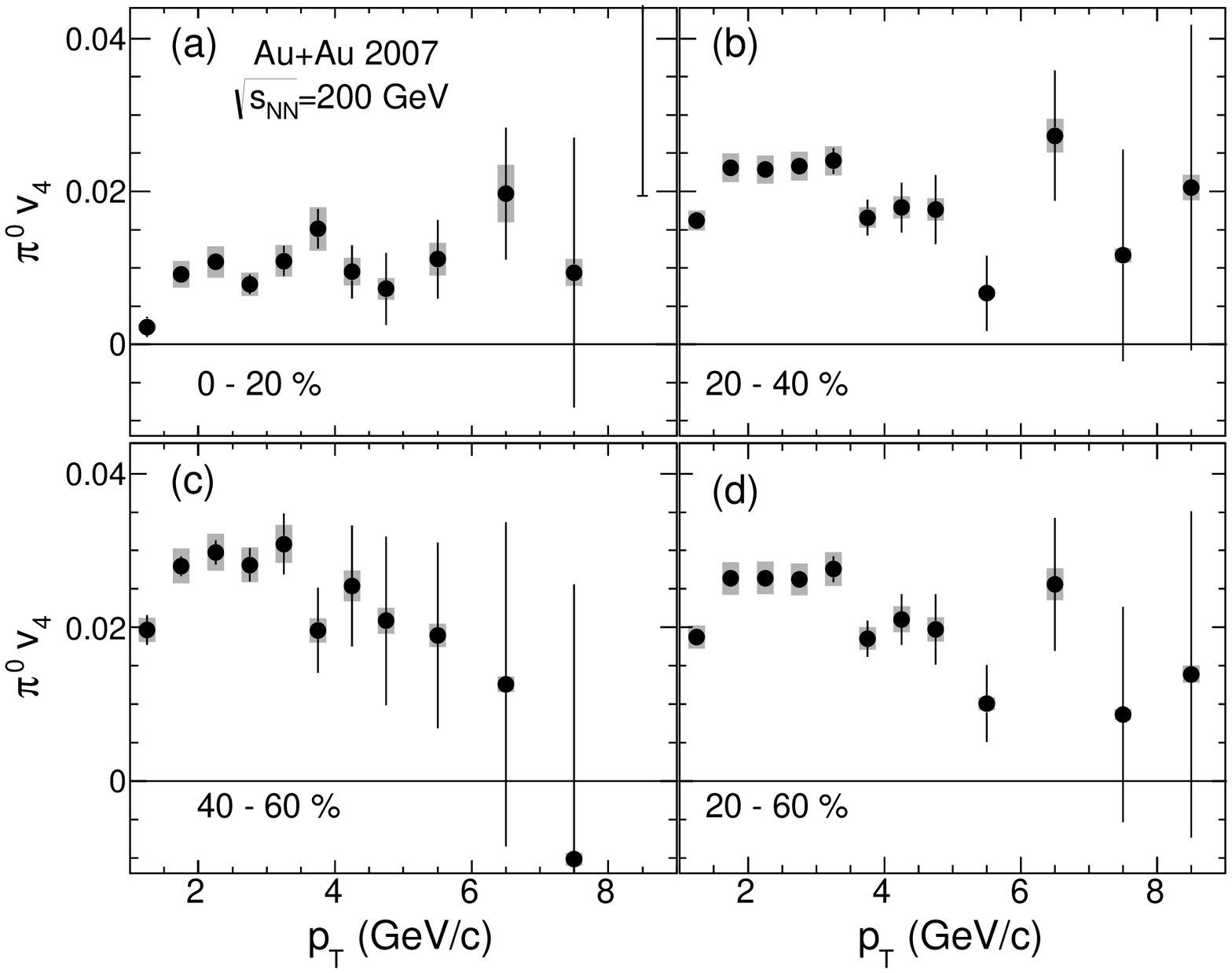}
\end{minipage}
\begin{minipage}{80mm}
\centering
\includegraphics[width=7.5cm]{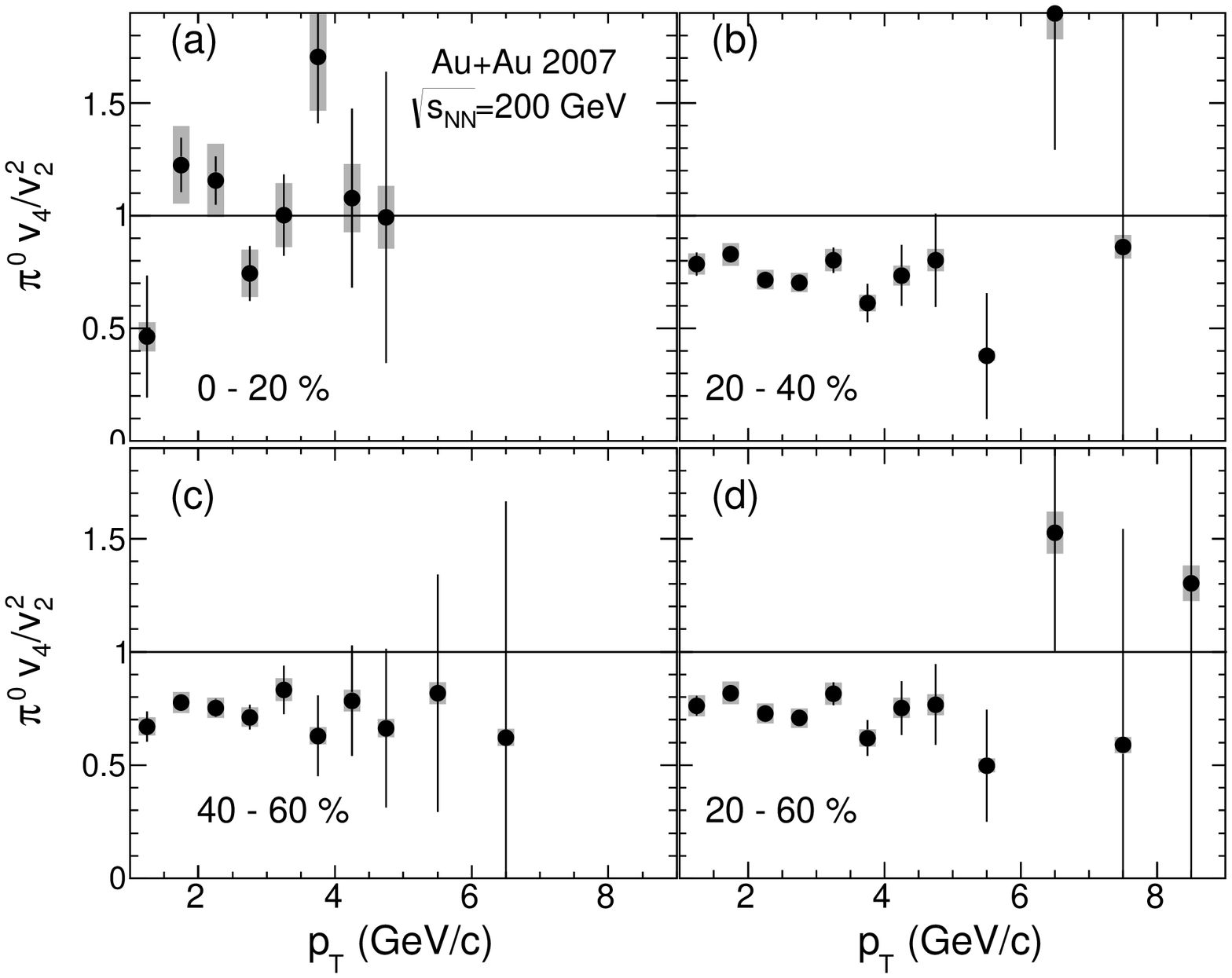}
\end{minipage}
\caption{(a, left) $v_4$ of $\pi^0$, and (b,right) $v_4/v_2^2$ for various centralities in 200GeV Au+Au collisions.}
\label{fig3}
\end{figure}

The transition from anisotropy driven by hydrodynamic flow to anisotropy 
driven by jet quenching can be probed by the ratio of $v_4/v_2^2$, where
$v_4$ is the fourth order and $v_2$ is the second order flow. Perfect fluid
hydrodynamics predicts a value of 0.5 for this ratio~\cite{Borghini:2005kd}.
The geometrical fluctuations and other dynamical fluctuations, as well as
viscous damping, can increase the magnitude of the ratio, especially in
central collisions~\cite{Gombeaud:2009ye}. At high $p_T$, the directions
that maximize collective flow and jet quenching may not be the
same~\cite{Jia:2012ez,Zhang:2012ha}. Therefore, this ratio could change
in the $p_T$ region where jet quenching begins to dominate. With the large
statistics, we were able to measure the $v_2$ and $v_4$ of $\pi^0$'s
with the same second-order event plane ($\Psi_2$) over a wide $p_T$ range.
Fig.~\ref{fig3}(a) shows the $v_4(p_T)$ of $\pi^0$ in 200GeV Au+Au
collisions~\cite{Adare:2013wop}.
We can see significant $v_4$ values even for $p_T>5$ GeV/$c$.
Fig.~\ref{fig3}(b) shows the $v_4/v_2^2$ ratios for $\pi^0$ obtained in
several centrality ranges~\cite{Adare:2013wop}. The ratios are approximately
independent of $p_T$, with values of $\sim$0.8-1.0 depending on centrality
selections. The constant ratios over the $p_T$ are not trivial at all given
several physics processes are involved, and may put additional constraint
on dynamical description of the medium.
The values for $p_T<\sim 5$ GeV/$c$ are consistent with our
prior observations of this ratio for inclusive charged hadron 
measurements~\cite{Adare:2010ux}.

\subsection{Path-length dependence of yield suppression}
\begin{figure}[h]
\centering
\includegraphics[width=14cm]{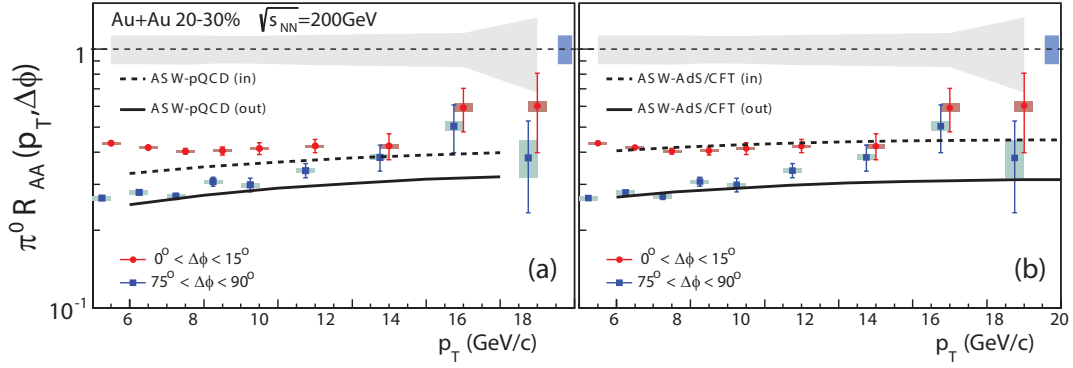}
\caption{$R_{\rm AA}(p_T,\Delta\phi)$ of $\pi^0$ in 20-30\,\% centrality for in-plane and out-of-plane. Data are compared with a pQCD-inspired model (left), and an AdS/CFT-inspired model (right).}
\label{fig4}
\end{figure}

Using the event plane information, we were also able to measure the
$R_{\rm AA}$ of $\pi^0$ as a function of $\Delta\phi$ with respect to the
event plane. Since the path length in the medium that partons traverse
changes by the emission angle with respect to the event plane (especially
in peripheral collision case), the angle dependence of the yields can be
associated with the path
length dependent energy loss of partons. We show the $R_{\rm AA}$ for in-
and out-of event planes for $\pi^0$s in 20-30\,\% central 200\,GeV Au+Au
collisions in Fig.~\ref{fig4}~\cite{Adare:2012wg}.
Depending on the energy loss models, the powers of the path-length
dependence change. The data favors an AdS/CFT-inspired (strongly coupled)
model rather than a pQCD-inspired (weakly coupled) model, implying that
the energy loss is $L^3$ dependent rather than $L^2$ dependence,
where $L$ denotes the path-length of partons in the medium.

\section{Fractional momentum loss of hadrons in A+A collisions}
\begin{figure}[h]
\begin{minipage}{50mm}
\centering
\includegraphics[width=5.0cm]{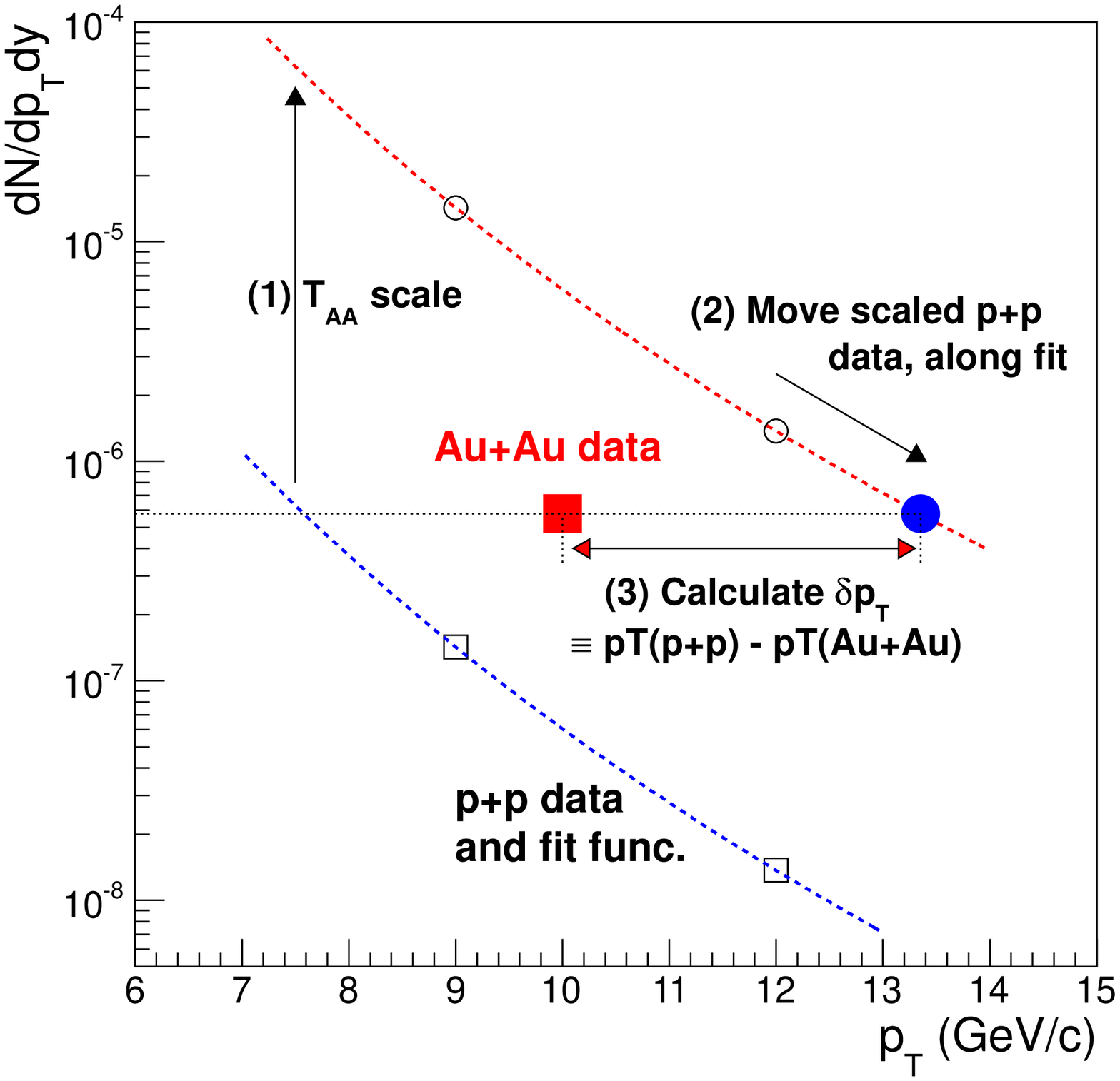}
\end{minipage}
\begin{minipage}{48mm}
\centering
\vspace{5mm}
\includegraphics[width=4.8cm]{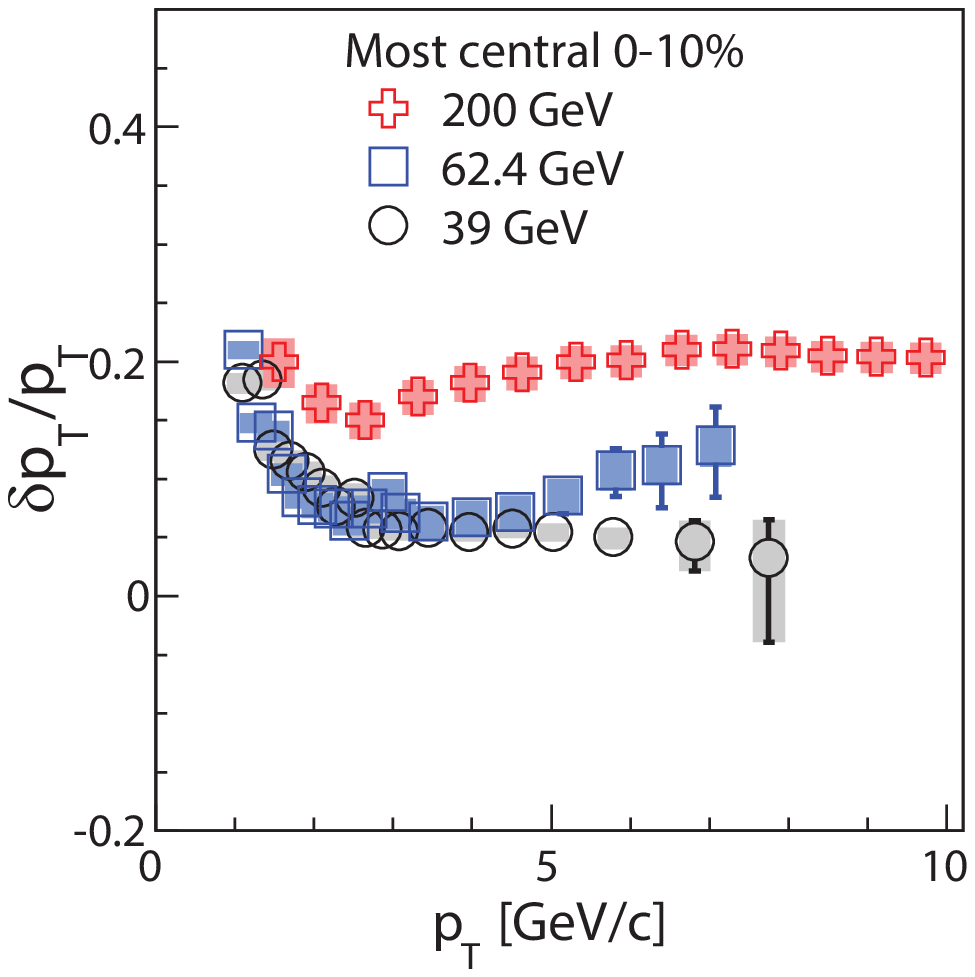}
\end{minipage}
\begin{minipage}{63mm}
\centering
\includegraphics[width=6.3cm]{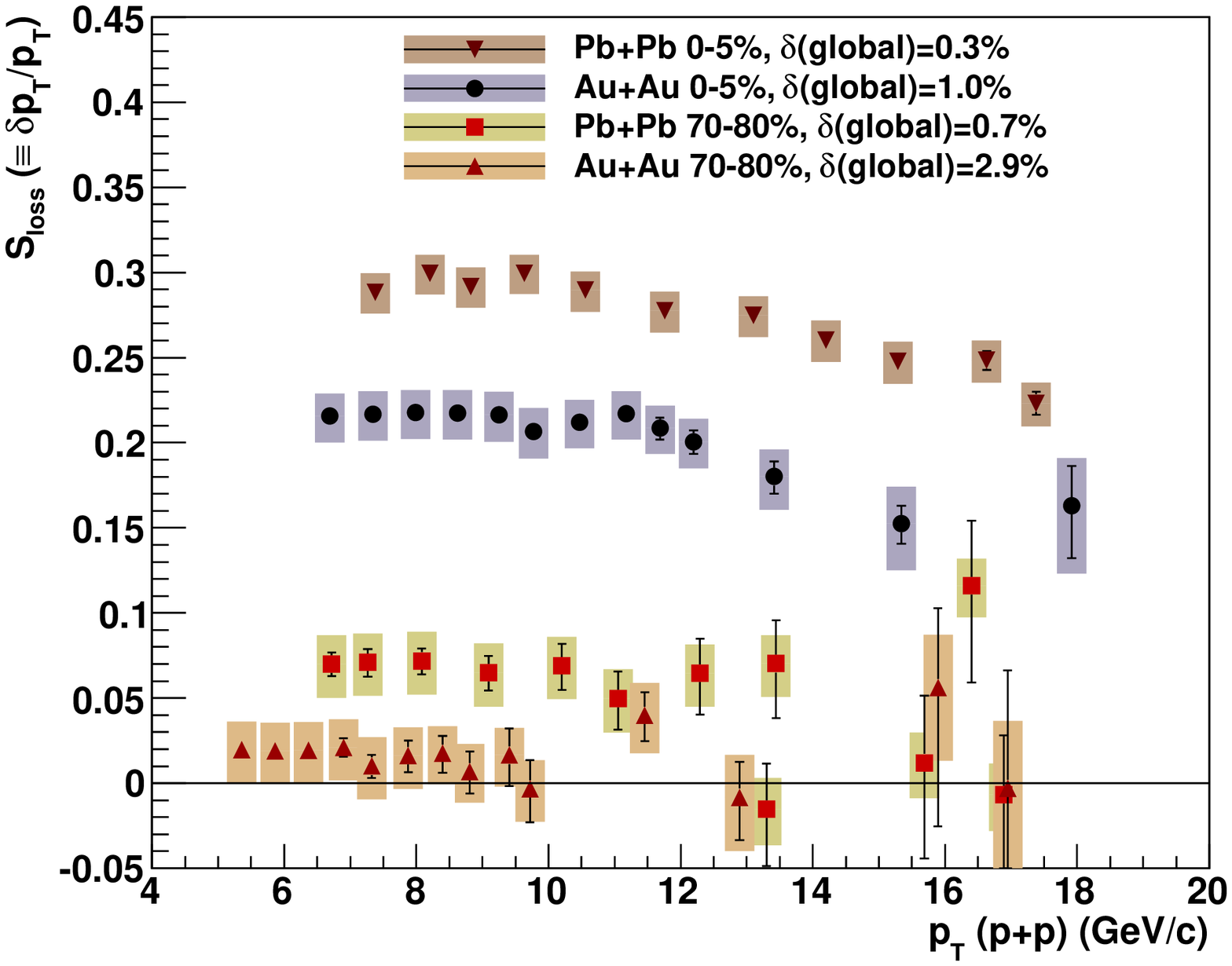}
\end{minipage}
\caption{(a, left) Method of calculating average $S_{\rm loss}$.
We scaled the $p+p$ yield by $T_{\rm AA}$ corresponding to centrality
selection of Au+Au data, shifted the $p+p$ points closest to Au+Au in yield,
and calculated momentum difference of $p+p$ and Au+Au points. (b, middle)
\sloss for $\pi^0$ for 0-10\,\% centrality 39, 62, and 200\,GeV
Au+Au collisions. (c, right) \sloss for $\pi^0$ in 200\,GeV Au+Au
collisions and charged hadrons in 2.76\,TeV Pb+Pb collisions.}
\label{fig5}
\end{figure}

\begin{figure}[h]
\centering
\includegraphics[width=9cm]{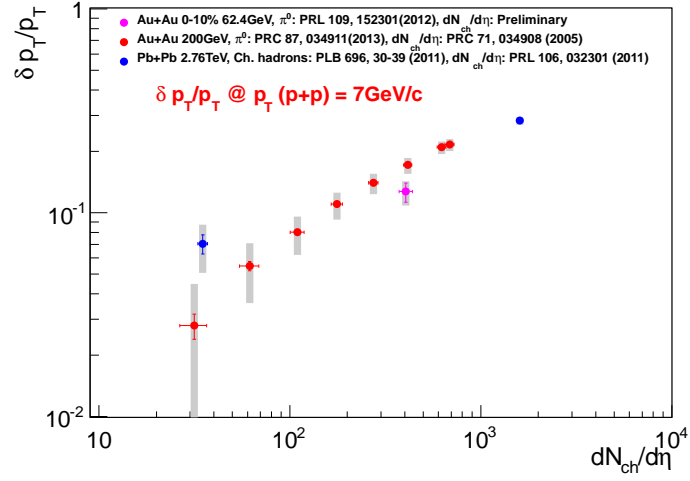}
\caption{$\delta p_T/p_T$ as a function of \dndeta for $\pi^0$ in 200\,GeV and 62.4\,GeV Au+Au collisions measured by PHENIX and charged hadrons in 2.76\,TeV Pb+Pb collisions measured by ALICE.}
\label{fig6}
\end{figure}

Experiments have been looking at the suppression of the yield at a given
$p_T$ to quantify the energy loss effect. However, the suppression is
primarily the consequence of the reduction of momentum of hadrons which
have exponential $p_T$ distributions. We have statistically extracted the
fractional momentum loss ($S_{\rm loss}\equiv \delta p_T/p_T$,
$\delta p_T \equiv p_T - p_T'$, where $p_T$ is the transverse
momentum of $p+p$ data, and $p_T'$ is that of Au+Au data) of the
partons using the hadron $p_T$ spectra measured in $p+p$ and Au+Au
collisions~\cite{Adare:2012wg}.
Fig.~\ref{fig5}(a) depicts the method to compute the $S_{\rm loss}$.
Using this method, we computed the $S_{\rm loss}$ in Au+Au collisions at
$\sqrt{s_{NN}}=$39, 62, and 200\,GeV as shown in
Fig.~\ref{fig5}(b)~\cite{Adare:2012uk}.
We also computed the $S_{\rm loss}$ in 2.76\,TeV Pb+Pb collisions using charged
hadron spectra measured by the ALICE experiment~\cite{Aamodt:2010jd}
as shown in Fig.~\ref{fig5}(c). $S_{\rm loss}$'s vary by a factor of six
from 39\,GeV Au+Au to 2.76\,TeV Pb+Pb collisions.
Naively, one expects that the energy loss is energy density dependent.
We plotted the $S_{\rm loss}$ against charged multiplicity, \dndeta,
at $p_T(p+p)=7$\,GeV/c, which is reasonably in hard scattering regime
as shown in Fig.~\ref{fig6}.
We assume \dndeta well represents the energy density of the system.
It is interesting to note that the trend of \sloss in Au+Au collisions
points to the most central points in Pb+Pb collisions at LHC. The 62\,GeV
Au+Au point and the most peripheral ALICE point are off the trend. These
features have not been found by looking at $R_{\rm AA}$'s. In order to
cross-check the new result, we have performed a power-law fit to the points
on $\delta p_T/p_T$ vs \dndeta, and compared the power with the
result obtained from a different method~\cite{Adare:2008qa}.
We fitted the points of this work with
$\delta p_T/p_T = \beta (dN_{\rm ch}/d\eta)^{\alpha}$ assuming
\dndeta$\propto$\Npart, and obtained $\alpha$ as 0.55$\pm$0.06.
Assuming the spectra shape is power-law with the power $n$, one can write
the relation between $S_{\rm loss}$ and \raa as:
\[ S_{\rm loss} \equiv \delta p_T/p_T = \beta N_{\rm part}^{\alpha}, \\
R_{\rm AA} = (1-S_{\rm loss})^{n-2} = (1-\beta N_{\rm part}^{\alpha})^{n-2} \]
Following this relation, we obtained the power $\alpha$ as 0.57$\pm$0.13 from
the fit to the integrated \raa as a function of \Npart in the
literature~\cite{Adare:2008qa}.
Thus, we confirmed that the powers obtained by two methods are very consistent.

\section{Summary}
We presented the recent results on high $p_T$ identified hadrons in Au+Au
collisions from the PHENIX experiment. The $R_{\rm AA}$ for
$\pi^0$ and $\eta$ are found to be very consistent. The second and
fourth order collective flow of $\pi^0$'s have been measured and found
that $v_4/v_2^2$ is consistent with the one observed in lower $p_T$ region,
which is not trivial given several physics processes are involved.
We found that the energy loss is $L^3$ dependent, where $L$ is the path
length of the partons in the medium. The $\delta p_T/p_T$'s of
high $p_T$ hadrons are computed from 39\,GeV Au+Au over to 2.76\,TeV Pb+Pb,
and found that they vary by a factor of six. We have seen a smooth trend
in $\delta p_T/p_T$ from RHIC energy to LHC energy when plotting against
charged multiplicity of the systems. We performed power-law fit to the
$\delta p_T/p_T$ vs \dndeta, and obtained a power that is very consistent
with the one obtained from the fitting to the integrated \raa.
We are going to add points from other systems to systematically
investigate the $\delta p_T/p_T$.


%
%
%






\begin{thebibliography}{99}
%
%
\bibitem{Wang:1998bha} 
  X.~-N.~Wang,
  Phys.\ Rev.\ C {\bf 58}, 2321 (1998).
\bibitem{Adler:2003qi} 
  S.~S.~Adler {\it et al.}  [PHENIX Collaboration],
  Phys.\ Rev.\ Lett.\  {\bf 91}, 072301 (2003).
\bibitem{Adcox:2003zm} 
  K.~Adcox {\it et al.}  [PHENIX Collaboration],
  Nucl.\ Instrum.\ Meth.\ A {\bf 499}, 469 (2003).
\bibitem{Adare:2010dc} 
  A.~Adare {\it et al.}  [PHENIX Collaboration],
  Phys.\ Rev.\ C {\bf 82}, 011902 (2010).
\bibitem{Adare:2012wg} 
  A.~Adare {\it et al.}  [PHENIX Collaboration],
  Phys.\ Rev.\ C {\bf 87}, 034911 (2013).
\bibitem{Borghini:2005kd} 
  N.~Borghini and J.~-Y.~Ollitrault,
  Phys.\ Lett.\ B {\bf 642}, 227 (2006).
\bibitem{Gombeaud:2009ye} 
  C.~Gombeaud and J.~-Y.~Ollitrault,
  Phys.\ Rev.\ C {\bf 81}, 014901 (2010).
\bibitem{Jia:2012ez} 
  J.~Jia,
  Phys.\ Rev.\ C {\bf 87}, 061901 (2013).
\bibitem{Zhang:2012ha} 
  X.~Zhang and J.~Liao,
  Phys.\  Rev.\  C 87, {\bf 044910} (2013).
\bibitem{Adare:2013wop} 
  A.~Adare {\it et al.}  [PHENIX Collaboration],
  Phys.\ Rev.\ C {\bf 88}, 064910 (2013).
\bibitem{Adare:2010ux} 
  A.~Adare {\it et al.}  [PHENIX Collaboration],
  Phys.\ Rev.\ Lett.\  {\bf 105}, 062301 (2010).
\bibitem{Adare:2012uk} 
  A.~Adare {\it et al.}  [PHENIX Collaboration],
  Phys.\ Rev.\ Lett.\  {\bf 109}, 152301 (2012).
\bibitem{Aamodt:2010jd} 
  K.~Aamodt {\it et al.}  [ALICE Collaboration],
  Phys.\ Lett.\ B {\bf 696}, 30 (2011).
\bibitem{Adare:2008qa} 
  A.~Adare {\it et al.}  [PHENIX Collaboration],
  Phys.\ Rev.\ Lett.\  {\bf 101}, 232301 (2008).
\end{thebibliography}







\end{document}